\documentclass[a4paper,fleqn,usenatbib]{mnras}

\usepackage{newtxtext,newtxmath}
\usepackage[T1]{fontenc}
\usepackage{ae,aecompl}

\usepackage{epsfig}
\usepackage{graphicx}	% Including figure files
\usepackage{amsmath}	% Advanced maths commands
\usepackage{amssymb}	% Extra maths symbols

\def\msun{{\rm M_{\odot}}}

\title [MOND or multiplicity?]
{The distribution of relative proper motions of wide binaries in GAIA DR2: MOND or multiplicity? }
\author[Clarke]{C.J. Clarke$^{1,}$\thanks{E-mail:cclarke@ast.cam.ac.uk}\\
$^1$ Institute of Astronomy, Madingley Rd, Cambridge, CB3 0HA, UK} 

\date{Submitted: }

\pagerange{\pageref{firstpage}--\pageref{lastpage}} \pubyear{2003}

\begin{document}
\def\lta{\mathrel{\spose{\lower 3pt\hbox{$\mathchar"218$}}
     \raise 2.0pt\hbox{$\mathchar"13C$}}}
\def\gta{\mathrel{\spose{\lower 3pt\hbox{$\mathchar"218$}}
     \raise 2.0pt\hbox{$\mathchar"13E$}}}
\def\Msun{{\rm M}_\odot}
\def\msun{{\rm M}_\odot}
\def\Rsun{{\rm R}_\odot}
\def\Lsun{{\rm L}_\odot}
\def\19{GRS~1915+105}
\label{firstpage}
\maketitle

\begin{abstract}
We examine the distribution of on-sky relative velocities for wide binaries previously assembled from
GAIA DR2 data and focus on the origin of the high velocity tail of apparently unbound systems which 
 may be interpreted as evidence for non-Newtonian gravity in the weak field limit. We
argue  that this tail is instead explicable in terms of a population of hidden triples, i.e. cases
where one of the components of the wide binary is itself a close binary unresolved in the GAIA data.
In this case the motion of the photocentre of the inner pair relative to its barycentre affects
the apparent relative proper motion of the wide pair and can make pairs that are in fact bound appear
to be unbound. We show that the general shape of the observed distributions can be reproduced using
simple observationally motivated assumptions about the population of hidden triples. 
\end{abstract}


\section {Introduction}
  It has long been recognised that ultra-wide binaries offer a potential
opportunity to scrutinise alternative gravity theories (see e.g. \citet{2012JPhCS.405a2018H}, %Hernandez et al 2012, \cite{2017IJMPD..2650067S}%Scarpa et al 2017
\citet{2018MNRAS.480.1778P}).% Pittordis \&  Sutherland 2018). 
This derives from
the fact that the gravitational acceleration experienced in solar
type binaries with separations of order $10^4$ AU is starting to become comparable
to that experienced in the outer reaches of spiral galaxies, where the
anomalously large circular velocity provides some of the most
convincing evidence for dark matter. Modified gravity theories 
(MOND, e.g. \citet{1983ApJ...270..365M}) %{1983ApJ...270..371M} %Milgrom 1983) 
instead
propose that the law of gravity is modified in the weak field limit.
A corollary of this assumption,  at least in the case of MOND-like theories
that posit a reduced role for the external field effect,  is  that wide binaries would be
subject to a larger gravitational acceleration, resulting in higher
orbital speeds, than in the case of purely Newtonian gravity.

 \citet{2019MNRAS.488.4740P} %Pittordis \& Sutherland (2019) 
have recently evaluated the evidence for
non-Newtonian gravity by examining the kinematics of a large
sample of candidate wide binary stars, identified in GAIA DR2 as being stellar pairs
with projected separation of $< 40000$  A.U., with consistent  parallax  
$< 5$ mas (corresponding to an estimated                   
distance of $< 200$pc) and with a relative velocity
component in the plane of the sky of  $< 3$ km s$^{-1}$. %The resulting sample
%contained  binaries with a parallax $< 5$ mas (corresponding to an estimated 
%distance of $< 200$pc); 
`Triples' (i.e. cases where there was evidence for additional
{\it resolved}  companions on a scale $> 100$ A.U.)  were eliminated from this sample. 
The normalised velocity (i.e. the on-sky relative velocity normalised
to that of a circular binary in the plane of the sky) was calculated for each pair
and the distribution of this quantity analysed over four bins of apparent separation
in the range $5000-20000$ A.U. 
 In a Newtonian scenario, this
distribution should be truncated at a  normalised velocity of
$\sqrt{2}$, this corresponding to a parabolic orbit in the plane of the
sky; pairs with higher normalised velocity would certainly be unbound
and would not be expected to be strongly represented in the observed population
since the timescale for the  change of separation of a parabolic binary
at a separation of a few kAU is short, $\sim$ a few $\times 10^5$ years.
Indeed, although Pittordis \& Sutherland were able to reproduce the distribution
of normalised velocities in a Newtonian scenario where hyperbolic  fly-bys generated
a significant high velocity tail, the required  stellar density far exceeded that
of the surveyed region. % (a volume of radius $200$ pc centred on the Sun).

  The key result of the analysis of \citet{2019MNRAS.488.4740P} %Pittordis  \& Sutherland (2019) 
is that the
distribution of normalised on-sky velocities, while being mainly concentrated at
values $< \sqrt{2}$, displayed a prominent `shelf' (flat high velocity tail)
extending up to the cut-off velocity for inclusion in the sample  (i.e. typically
up to a normalised velocity of $\sim 7$). Pittordis \& Sutherland  also explored explanations 
for this high velocity `shelf' in terms of a variety of  alternative gravity theories,  finding that while MOND theories without external field
effect are incompatible with the data, there is still a potential role for
MOND with an external field effect, although the data are not decisive in
this regard.

 Here we put forward an alternative hypothesis, i.e. that this `shelf' is a consequence
of the higher order multiplicity of the wide binary sample.
 Note that the possible influence of hidden multiples
has previously been discussed by \cite{2012JPhCS.405a2018H} %Hernandez et al 2012 
and also
by \citet{2019MNRAS.488.4740P} %Pittordis \& Sutherland 2019        
although  this has not been quantified in
detail. 
In the GAIA data the `wide' pair is well resolved and, as noted above,  systems with 
evidence of further companions
on a scale $> 0.5$ mas  (typically $100$ A.U.) were eliminated from the sample. If the system however
contains an {\it unresolved} inner binary of non-unit mass ratio, then there is an additional
shift of the photocentre of this inner binary with respect to  its centre of mass. Consequently this would introduce
an error into the  derived relative proper of the inner pair and its wide companion. In this
paper we generate a synthetic population of wide binaries, including a subset with unresolved
components, and evaluate how the shift between the motion of the photocentre and the barycentre of
the inner pair can distort the measurements of relative proper motions within wide
binaries. Note that in our
analysis we only consider inner components with a restricted range of separations: close
enough that they are unresolved by GAIA (i.e. $< 100$ 
 A.U.) and wide enough that they would not
have been eliminated from the sample  on the basis of an unacceptable astrometric solution.
(Note that inner binaries with separations  $<$ a few AU would manifest a variable
proper motion over the $22$ month baseline of the GAIA DR2 release and would not have been
included in the wide sample analysed here).

 In Section 2 we present the details of our simulations which use observationally motivated
parameters in order to synthesise the effect of  higher order multiplicity. Our simulations readily produce a high
velocity `shelf' that is very similar to that observed in the GAIA DR2 data and thus suggest that the
observed velocity distribution should {\it not}  be regarded as {\it prima facie} evidence for non-
Newtonian gravity. We point out that with a more exact definition of the separation limits for
the inner binaries that are hidden in the DR2 sample, the observed distribution of normalised
velocities instead provides a  tool for characterising higher order multiplicity within wide solar
type binaries.

\section {The effect  of higher order multiplicity on the normalised velocity distribution}

\subsection {Simulation method}
  In order to assess the effect of concealed inner binaries on wide binary kinematics, we construct
a fiducial model, focusing on wide binaries with a solar type primary and separations on the sky in the four logarithmically spaced bins considered by \citet{2019MNRAS.488.4740P} %Pittordis \& Sutherland 2019 
in the range  $5000-20000$ A.U., using the inter-bin variation in the
numbers of binaries  to assign the probability distribution within each bin. 
The
components of the wide binaries are selected at random from a distribution with mass ratio,
$q$,  where the
fraction of systems per unit interval of mass ratio scales as $q^{-1.1}$ \citep{2017ApJS..230...15M};%{ApJS..230...15M};%(Moe \& di Stefano 2017);i
given the magnitude limits imposed by Pittordis
\& Sutherland we only consider outer pairs with a mass ratio
in the range $0.5-1$. 
Such binaries are uniformly distributed in cos $i$ (where $i$ is the angle between the line of sight and
the normal to the binary orbit). The binary eccentricities are selected from a thermal distribution
(fraction per unit interval of eccentricity, $e$, scaling as $e$ \citep{1975MNRAS.173..729H}) % Heggie 1975) 
and the relative velocity between
the components of the wide pair in the plane of the sky is calculated. As expected, this gives rise
to a distribution that truncates at normalised velocity ratio of $\sqrt{2}$ and matches the
distributions in the Newtonian case shown in \citet{2019MNRAS.488.4740P}. %Pittordis \& Sutherland (2019).

 We then furthermore assume that a fraction $f_{triple}$ of these wide pairs have a
primary or secondary component that is itself a binary with separation $a$, that is
uniformly distributed  in log separation over the range $a_{min}$ to $a_{max}$. As noted above,
$a_{min}$ has to be large enough that the astrometric solution of the component
was not flagged  as exhibiting  variability over the duration of the GAIA DR2
experiment. $a_{max}$ has to be small enough that the inner pair was not resolved
by GAIA and rejected as a `triple' from the sample. We here adopt $a_{min} = 3$ A.U.
and $a_{max}= 100$ A.U. based on the $22$ month duration of the GAIA DR2 dataset
and the resolution limits for the sample quoted by \citet{2019MNRAS.488.4740P}. %Pittordis \& Sutherland (2019). 
We assume 
that the orbital plane of each inner binary is  randomly inclined with respect
to the plane of the outer binary and that the mass ratio and eccentricity distribution
are randomly selected from the same distributions given above for generating
the outer pairs, with the mass ratio distribution extending down to $q=0.1$.. {\footnote{ There is some evidence that the mass ratio distribution becomes more
weighted towards higher $q$ for small separations, although the errorbars on mass ratio distribution
slopes are large and in any case are not evaluated specifically for the inner pairs of multiples.
In order to investigate the sensitivity of the resulting velocity distributions on
the assumed mass ratio distribution we have also considered, for inner binaries, the mass ratio distribution quoted
by Moe \& di Stefano for binaries with separations around 1 AU: these  are parameterised with power law slopes of
$-0.1$ and $-0.5$ in the domains respectively below and above $q=0.3$. This change slightly boosts
the perturbations associated with hidden triple components as it gives more
prominence to pairs of intermediate ($q \sim 0.6 \pm 0.3 $)  for which the photocentre-barycentre motion is maximal; the effect is however small ($< 10 \%$ in average
normalised photocentre shift). We thus conclude that the results are robust against plausible
variations in the assumed mass ratio distribution.}}
The components of the inner binary are then assigned G band magnitudes using the
mass luminosity relationship given by equation (3) of  \citet{2019MNRAS.488.4740P}, %Pittordis \& Sutherland (2019),
and from this the on sky motion of the inner binary's photocentre relative to its barycentre
is calculated. For such hidden triples, the relative proper motion of the two wide components is 
evaluated 
as the velocity  in the sky plane of the inner binary's photocentre 
with respect to its distant companion. Binaries and hidden triples are then treated
identically {\footnote{ We have also considered the small  modification associated with
the fact that the observer who cannot resolve the inner pair will under-estimate its mass
from the mass-luminosity relation:  
%in the
%case of the hidden triple population is that one should also take account of the fact that the
%observer will slightly underestimate the mass of the unresolved pair based on its total
%luminosity: 
this however only changes the circular velocity assigned by of order $1 \%$. }} in that the normalised velocity (i.e. the ratio of the  on-sky relative motion to
that of a circular binary in the plane of the sky) is calculated for each wide system,  
 thus
generating a
distribution over the synthesied population.
We adopt $f_{triple}=0.5$, motivated by the results  of 
\citet{2015ApJ...799....4R} who found that in a sample of $212$ wide binaries, $100$ had additional
inner components with separations in the range of a few to a few hundred A.U. (see also \citet{2014AJ....147...87T} 
%Tokovinin 2014 and Halbwachs et al 2017 
and \citet{2017MNRAS.464.4966H} for evidence of a high inner multiple fraction within wide
pairs). 

\subsection {Simulated results}
  
Figure 1 presents the results of the above described fiducial
model in each of the separation bins considered by Pittordis \&
Sutherland. We compare the predictions of this model for
the distribution of normalised velocities (dashed) to the 
corresponding distributions observed in each bin (solid), using the sub-samples of the GAIA DR2 data 
for which  the relative velocity error is less
than $0.25$ times the circular velocity. The number of synthesised binaries
matches the sample number in each separation bin 
(respectively $629,428,270$ and
$134$). 
%of normalised : in each bin the simulated data contains the l algorithm  applied to $955$  synthetic pairs (solid histogram) compared (dashed histogram) 
% with the corresponding observed distribution of normalised velocities for binaries
%in the range $5000$ to $7000$ A.U. from the GAIA DR2 analysis of \citet{2019MNRAS.488.4740P}. %Pittordis \& Sutherland (2019). 
It can
be seen that the simulated populations recover the main features of the observed distribution in all separation bins without any fine tuning of the models,
a point that is emphasised by the excellent agreement between the
corresponding cumulative distributions shown in Figure 2 (we have checked
that the small deviations between the synthesised and observed
cumulative distributions seen in Figure 2 are not statistically
significant given the sample sizes involved). 

 While the majority
of the population has normalised velocity $< \sqrt{2}$ there is a significant tail to higher velocities which
is entirely composed of hidden triple systems where the relative motion of the photocentre and barycentre
of the unresolved inner pair is sufficient to scatter bound systems so that they appear to be unbound in this
plot. It can readily be seen how the magnitude of relative motion between  the barycentre and photocentre of an
unresolved pair
can affect the apparent kinematics of wide binaries.  For example, the relative velocity of a solar mass
circular binary of separation $10$ A.U. is around $10$ km s$^{-1}$; if the internal mass ratio of this
binary were  $0.25$, for example, the relative velocity between the barycentre and photocentre  would be  $\sim  2$ km s$^{-1}$ which is
a significant value compared with the circular velocity of the wide ($\sim 5000$ A.U.) binary
($\sim 0.5$ km s$^{-1}$).  The featureless nature of this  excess at high normalised velocities
stems  from the combined effects of a broad distribution of mutual inclinations, ecentricities and mass
ratios of the inner pair.

%\begin{figure}
%\begin{center}
%\includegraphics[width=\columnwidth,angle=0]{binvf1.pdf}
%\caption{Normalised velocity distribution for the fiducial model
%(solid) compared with data in the $5000-7000$ A.U. range (dashed) from
%Pittordis \& Sutherland (2019)}
%\end{center}
%\end{figure}

%\begin{figure*}
%    \centering
%    \begin{subfigure}{0.49\linewidth}
%        \centering
%        \includegraphics[width=\linewidth]{5to750logslopeha.pdf}
%        \caption{Initially large ($R_C = 100~\mathrm{AU}$) disc.}
%        \label{fig:low_high100}
%    \end{subfigure}
%    \begin{subfigure}{0.49\linewidth}
%        \centering
%        \includegraphics[width=\linewidth]{7to1050logslopeha.pdf}
%        \caption{Initially small ($R_C = 10~\mathrm{AU}$) disc.}
%        \label{fig:low_high10}
%    \end{subfigure}
%     \addtocounter{figure}{-1}}
%     \end{figure}

%\begin{figure*}
%\begin{center}
%\includegraphics[width=\columnwidth]{5to750logslopeha.pdf}
%\includegraphics[width=\columnwidth]{7to1050logslopeha.pdf}\\
%\includegraphics[width=\columnwidth]{10to1450logslopeha.pdf}
%\end{center} 
%\end{figure}

\begin{figure}
\begin{center}
\begin{minipage}[]{8.8cm}
\includegraphics[width=4.4cm]{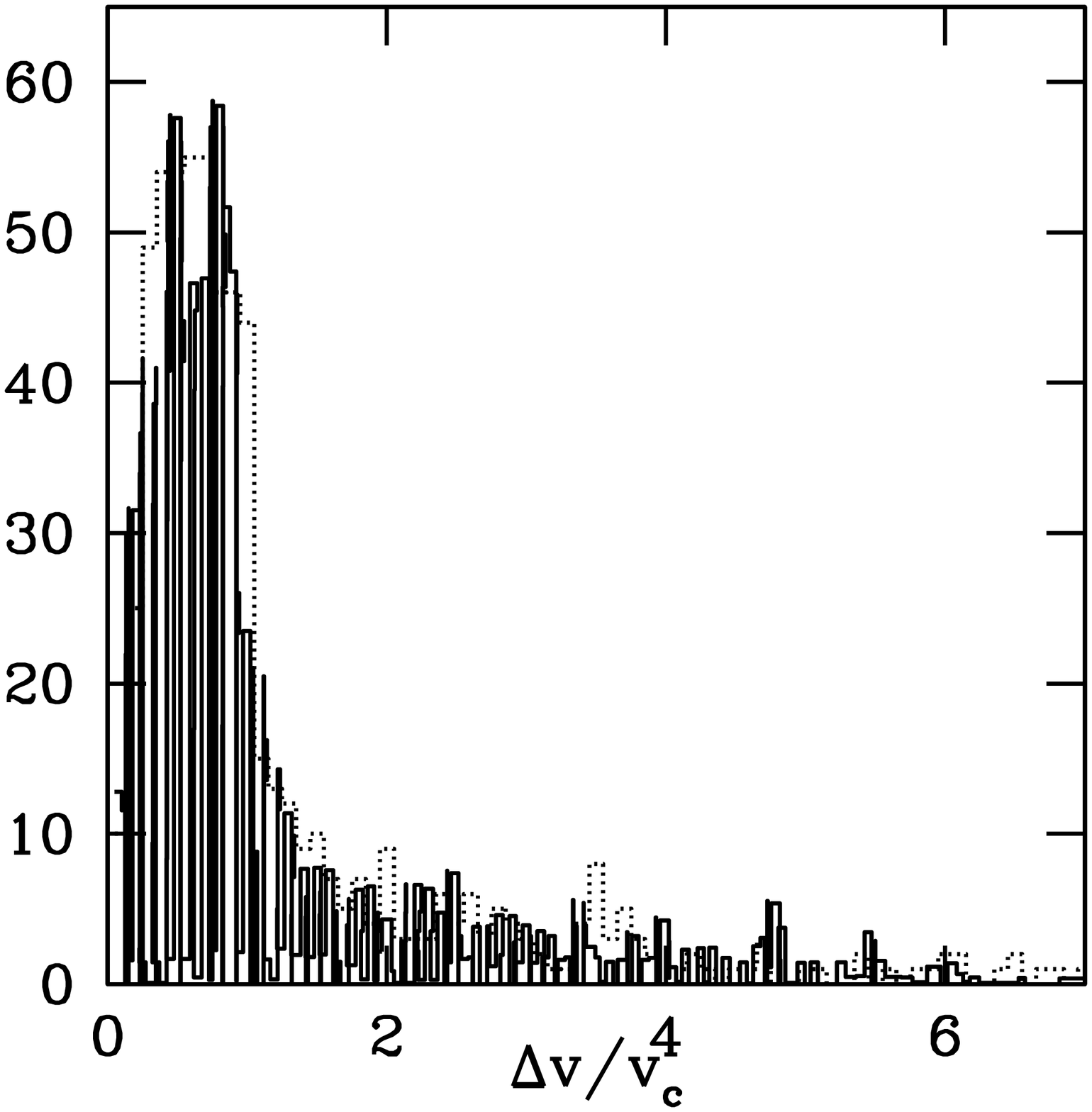}
\includegraphics[width=4.4cm]{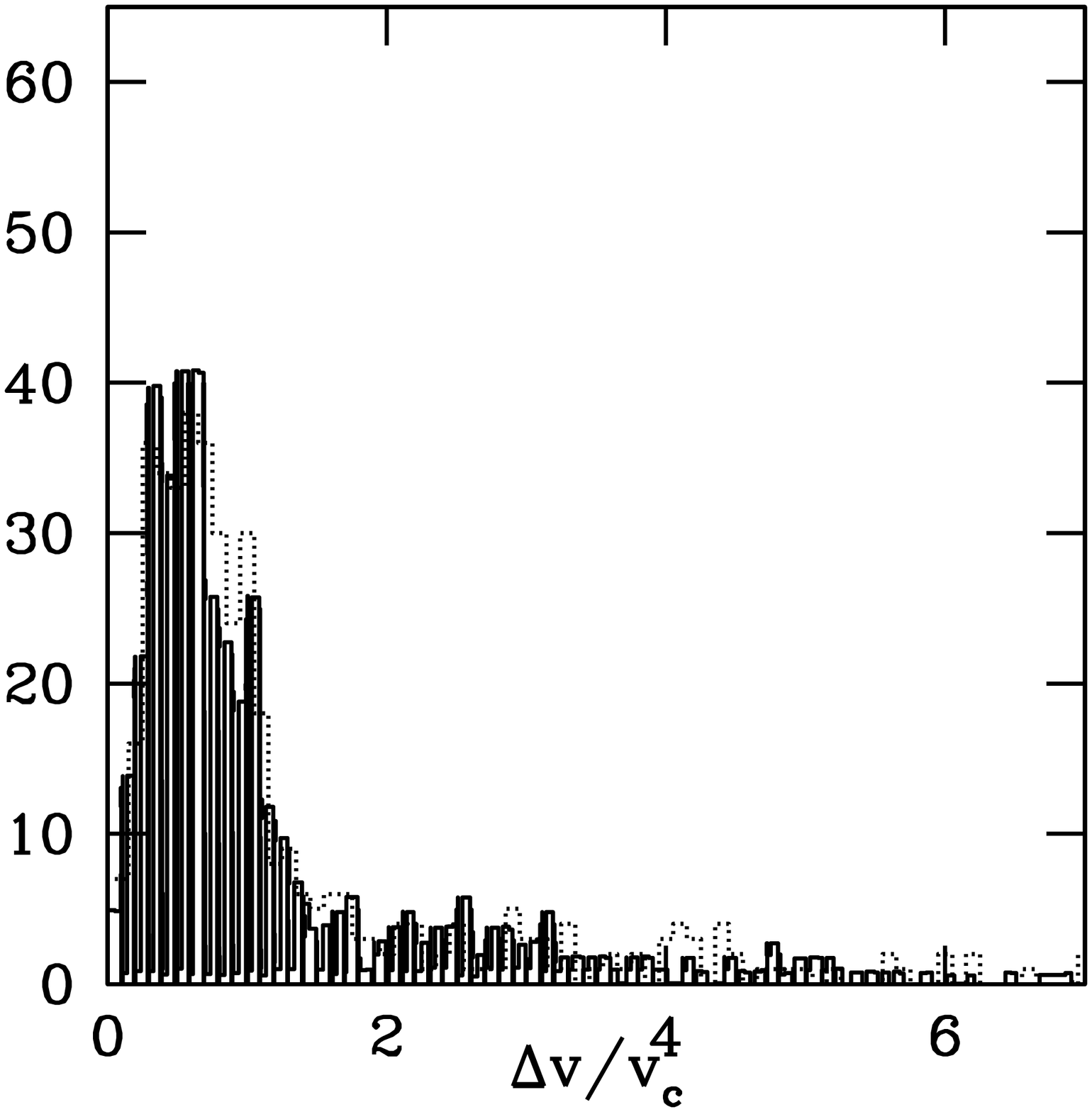}
\includegraphics[width=4.4cm]{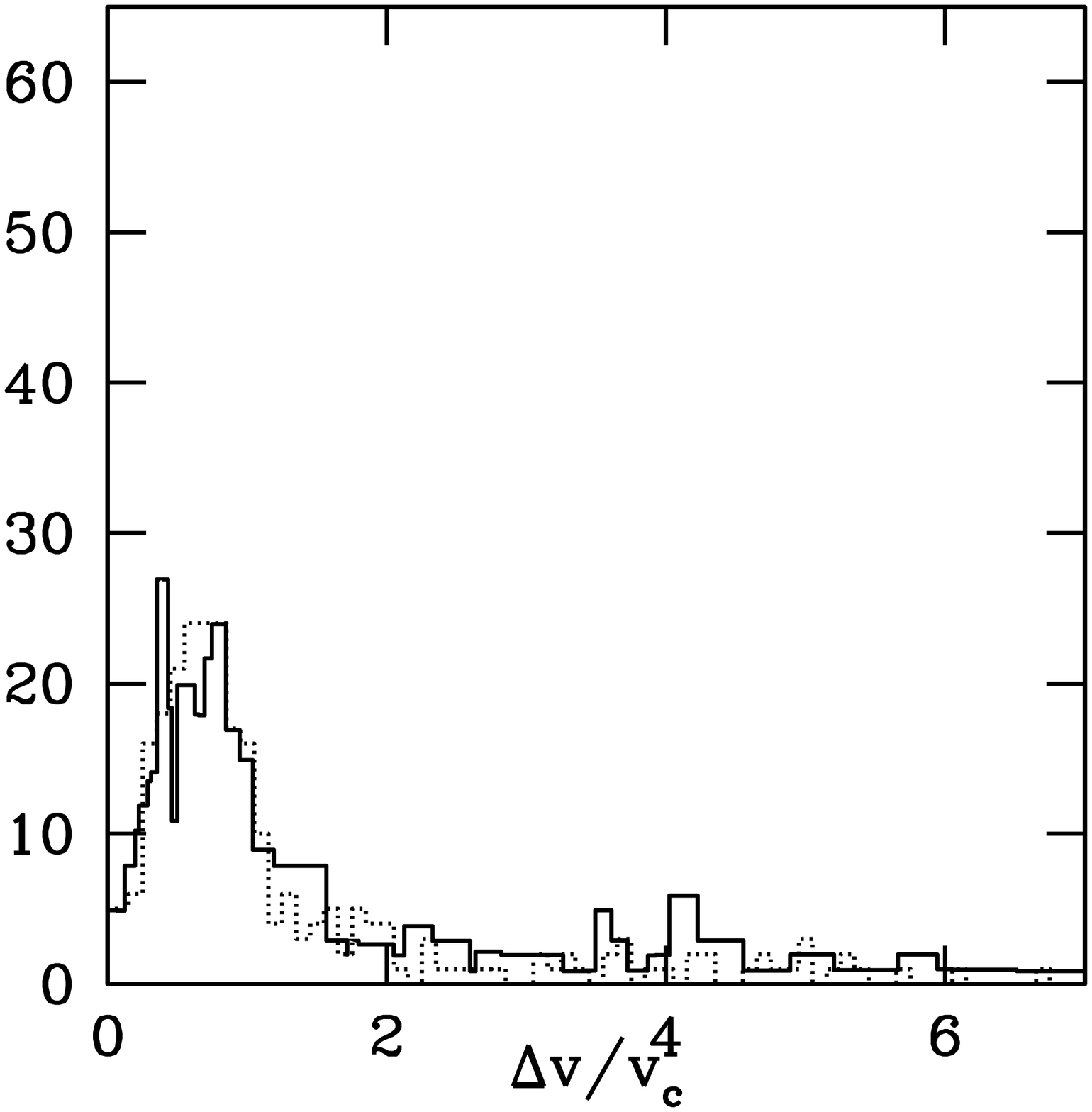}
\includegraphics[width=4.4cm]{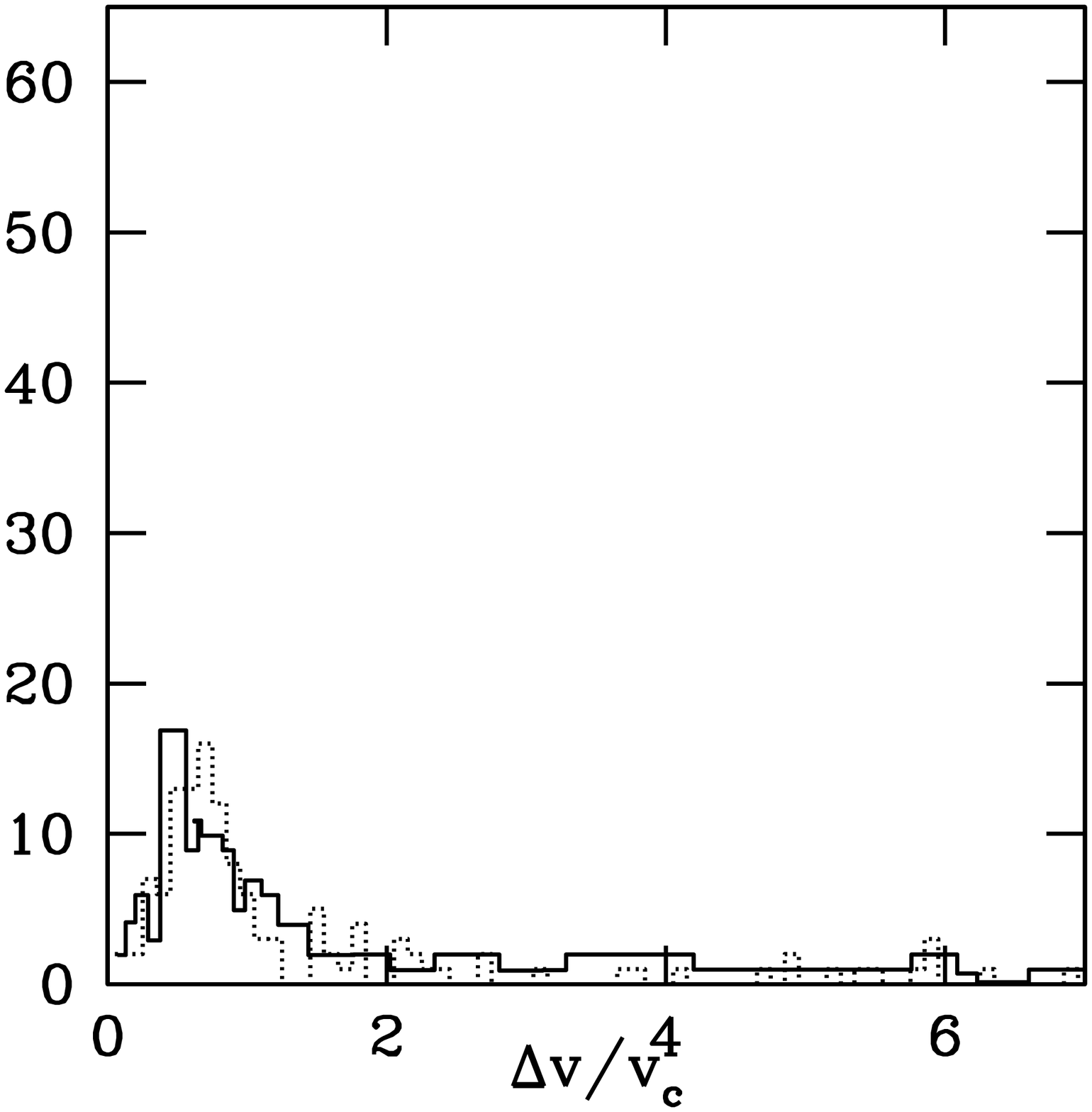}
\end{minipage}
\caption{Normalised velocity distribution for the fiducial model (dashed: see Section 2.1)
compared with data (solid) in the separation ranges:  5000- 7000 A.U. (upper left), 7000-10000 A.U. ( upper right), 10000-14000 A.U. (lower left) and 14000-20000 A.U. (lower right)  from \citet{2019MNRAS.488.4740P}.} %Pittordis
%\& Sutherland (2019).}
\end{center}
\end{figure}

 We also note the consistency between the model predictions and observations
concerning trends with increasing binary separation. In the low error
sub-samples analysed in Pittordis \& Sutherland, the number of 
systems with normalised velocity ratio $> \sqrt{2}$ is 
$159$ in the separation range $5000-7000$ A.U. and 
$53$ in the separation range $14000-20000$ A.U.. As the authors
note, an explanation in terms of flybys would instead imply an
increase in the absolute  number of such objects at large separation. In the
framework of hidden triples, however, the velocity component
associated with a given population of close pairs would represent
an increasing fraction of the circular velocity of the distant
pair at large separations and so would imply that the {\it fraction}
of wide pairs with high normalised velocity ratios should increase
at large separations. Observationally this is indeed the case
(e.g. $159/629 \sim 0.25 $ pairs in the $5000-7000$ A.U. range have normalised
velocities $> \sqrt{2}$ compared with the corresponding fraction
$53/134 \sim 0.4$ in the $5000-7000$ A.U. range.)   

 \begin{figure}
\begin{center}
\begin{minipage}[]{8.8cm}
\includegraphics[width=4.4cm]{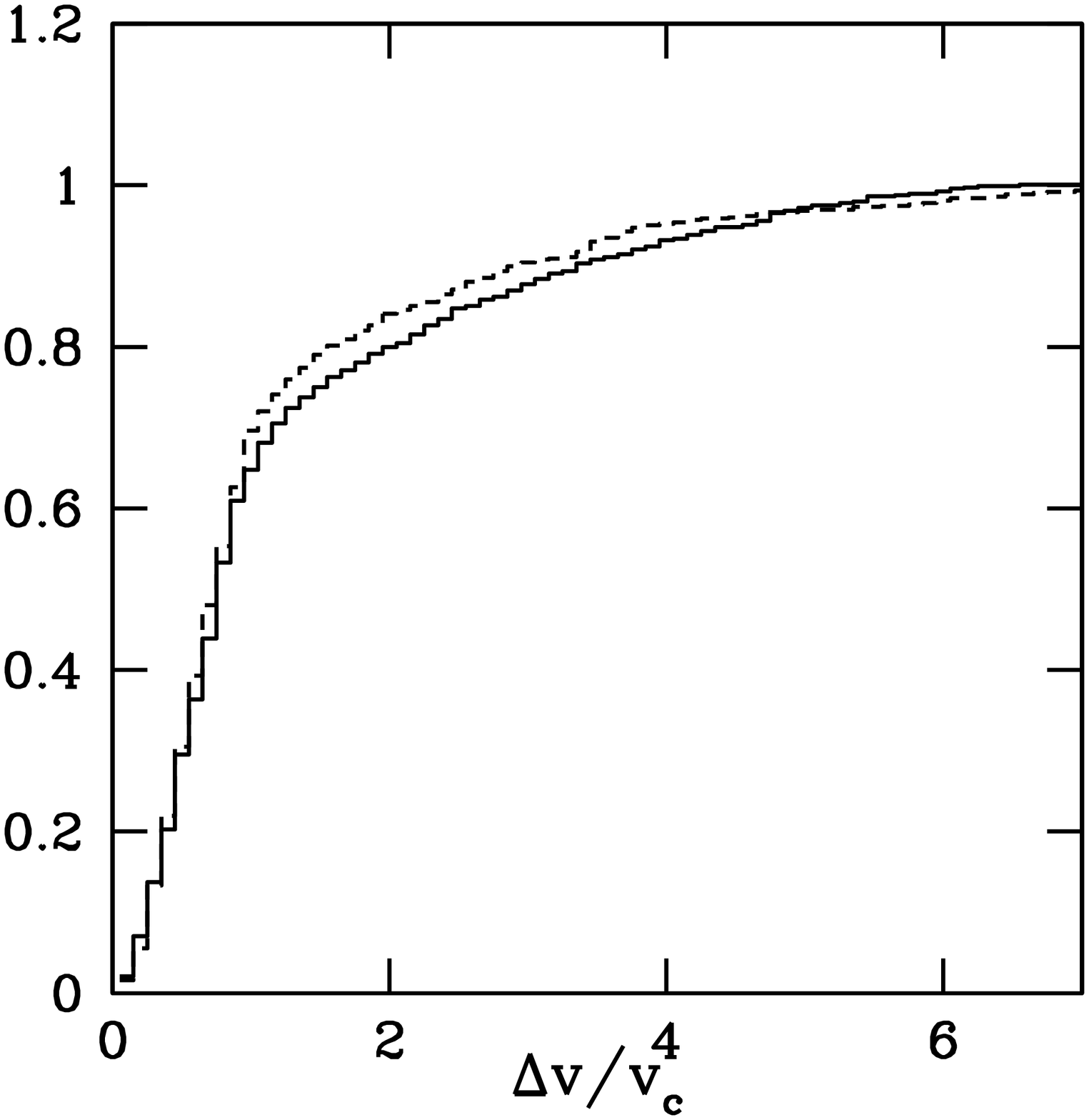}
\includegraphics[width=4.4cm]{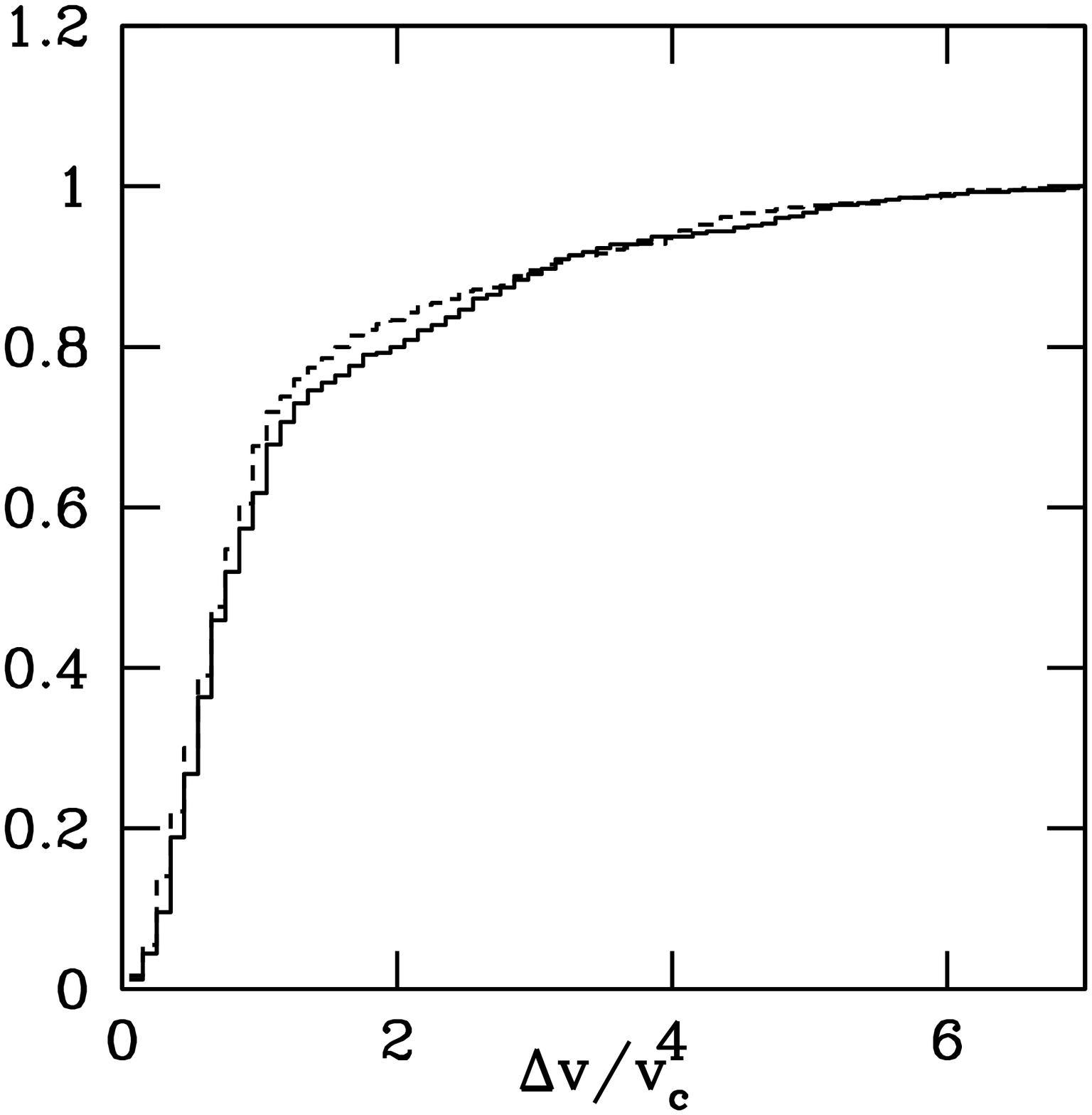}
\includegraphics[width=4.4cm]{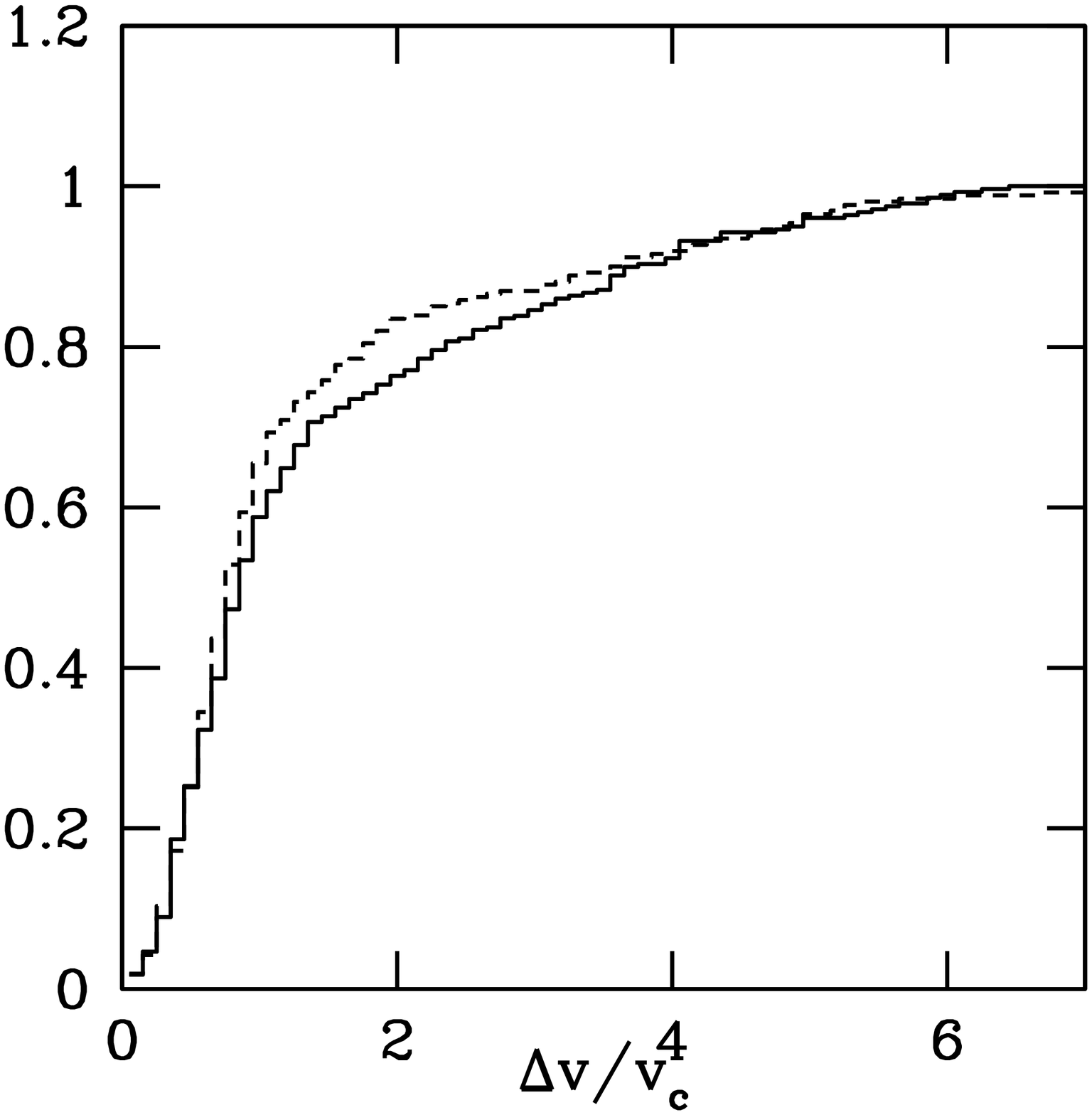}
\includegraphics[width=4.4cm]{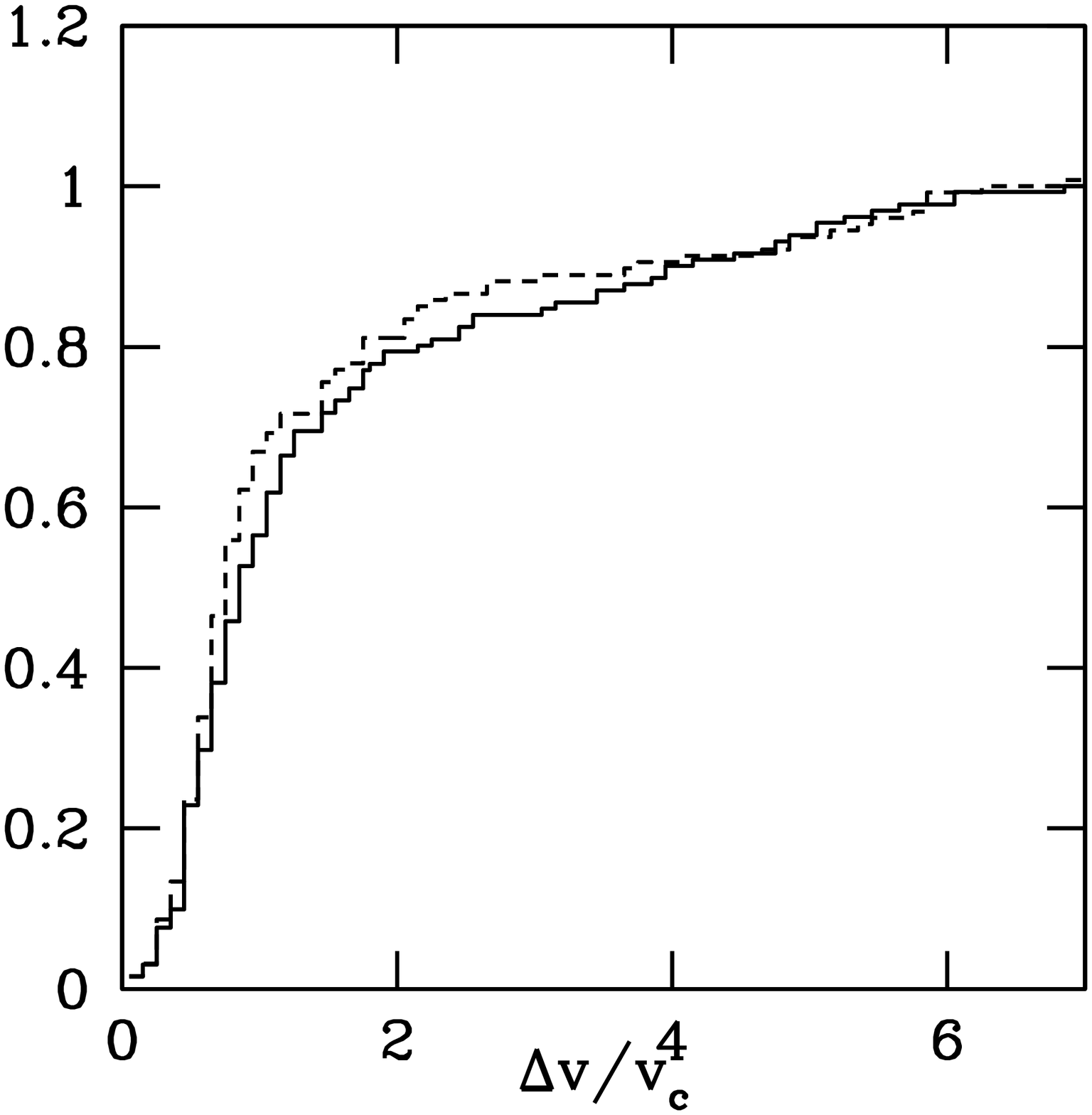}
\end{minipage}
\caption{Cumulative distributions corresponding to histograms shown in Figure 1.}
\end{center}
\end{figure}
\section {Conclusions}

  We have shown that the observed population of wide binaries with normalised velocities in excess
of $\sqrt {2}$  as derived from GAIA DR2 data is readily explicable in terms of contamination of the sample
by unresolved close ( a few to $\sim 100$ A.U.) inner multiple components. Such `hidden triples' introduce
an error in the kinematics of the wide pair which stems from the relative motion of the barycentre and
photocentre of the inner pair. In consequence some wide systems can appear to be unbound  when in
fact they are bound hierarchical multiple systems governed by Newtonian dynamics.

 We find that the `shelf' of high velocity ratios can be  entirely
attributable to hidden triples if the triple fraction is as high
($\sim 50 \%$) as suggested by local surveys \citep{2015ApJ...799....4R}.%(Riddle et al 2015). 
This
is clearly a hypothesis that can be tested by targeted investigations
(e.g. high contrast adaptive optics imaging) of objects with anomalously high velocity ratios. Likewise the hidden triple hypothesis would imply
possible  anomalies in the locations of stars in the colour magnitude diagram
(e.g. \citet{2018ApJ...857..114W})%Widmark et al 2018). 
Hidden triples can also  
generate discrepant
proper motions over longer time baselines.  We note that our fiducial
model predicts that, when comparing proper motions of wide binaries obtained
from the ten year Hipparcos survey and the $22$ month DR2 survey,
 $<10 \%$ of the sample would be expected to show discordant proper motions
at a level comparable with typical   Hipparcos errors. Such
a figure is compatible with the comparison presented in
\citet{2019IJMPD..2850101H}. %Hernandez et al 2019. 
 
The important contribution from hidden triples raises a caution against interpreting this population of apparently unbound pairs as evidence
for a modification of the gravitational acceleration in the weak gravity regime (i.e. as evidence for
MOND or a variant theory). At the very least, further
examination of this subject will need to examine a subset of proper motion data where there are independent
limits on the existence of inner binary components. Conversely, if retaining the assumption of
Newtonian gravity, GAIA data has the possibility to provide statistical constraints on higher
order multiplicity on samples that are far larger than those assessed in conventional
multiplicity surveys.

\section{Acknowledgments}
 I am grateful to useful input from Will Sutherland and to the referee for
comments which have improved the paper.

\bibliographystyle{mnras}
\bibliography{mybib} % if your bibtex file is called example.bib

\end{document}